\pdfoutput=0
\documentclass[12pt]{iopart}

\usepackage{graphicx,setspace}
\usepackage{psfrag}
\usepackage{amsfonts}
\expandafter\let\csname equation*\endcsname\relax
\expandafter\let\csname endequation*\endcsname\relax
\usepackage{amsmath}
\usepackage{amssymb, amsthm}
\usepackage{mathrsfs}
\usepackage{epstopdf}
\usepackage{float}
\usepackage{color}
\usepackage{subfigure}
\usepackage{indentfirst}
\usepackage{array}
\usepackage{bm}
\begin{document}

\newtheorem{definition}{Definition}
\newtheorem{lemma}{Lemma}
\newtheorem{remark}{Remark}
\newtheorem{theorem}{Theorem}
\newtheorem{proposition}{Proposition}
\newtheorem{assumption}{Assumption}
\newtheorem{example}{Example}
\newtheorem{corollary}{Corollary}
\def\ep{\varepsilon}
\def\Rn{\mathbb{R}^{n}}
\def\Rm{\mathbb{R}^{m}}
\def\E{\mathbb{E}}
\def\hte{\hat\theta}

\title[The most probable dynamics of receptor-ligand binding on cell membrane]{The most probable dynamics of receptor-ligand binding on cell membrane}

\author{Xi Chen$^{1}$, Hui Wang$^{2,*}$,  Jinqiao Duan$^{3}$}

\address{$^1$ School of Statistics, Xi 'an University of Finance and Economics, Xi'an 710100,China}
\address{$^2$School of Mathematics and Statistics, Zhengzhou University, Zhengzhou 450001, China}
\address{$^3$ Department of Applied Mathematics, Illinois Institute of Technology, Chicago, IL 60616, USA}
\ead{huiwang2018@zzu.edu.cn}
\vspace{10pt}
\begin{indented}
\item[]February 2023
\end{indented}

\begin{abstract}
We devise a method for predicting certain receptor-ligand binding behaviors, based on stochastic dynamical modelling. We consider the dynamics of a receptor binding to a ligand on the cell membrane, where the receptor and ligand perform different motions and are thus modeled by stochastic differential equations with Gaussian noise or non-Gaussian noise. We use neural networks based on Onsager-Machlup function  to compute the probability $P_1$ of the unbounded receptor diffusing to the cell membrane. Meanwhile, we compute the probability  $P_2$ of extracellular ligand arriving at  the cell membrane by solving the associated Fokker-Planck equation. Then, we could predict the most probable binding probability by combining $P_1$ and $P_2$. In this way, we conclude with  some indication about where the ligand will most probably encounter the receptor, contributing to better understanding of cell's response to external stimuli and communication with other cells.
\end{abstract}

\vspace{2pc}
\noindent{\it Keywords\ } {binding probability; most probable transition pathway; Onsager-Machlup function; Fokker-Planck equation}
%
%
%
%

\section{Introduction}

There has been an explosion of interest in the effect of noise in cell biology in recent years \cite{Bressloff2014,Schuss,Bialek,Lauffenburger,Qian,Anderson}. Stochastic cell biology has attracted the attention of many researchers in biophysics, statistical physics and stochastic dynamics \cite{Logan}.  As the cell is an open system that exchanges energy and matter with the environment, the molecules inside and outside the cell keep moving  by different transport mechanisms. Besides, the  cells are in highly complex and noisy environment, so the motions are stochastic \cite{Bressloff2014}.

Many essential cell functions, such as monitoring the environment, responding to external stimuli and communicating with neighboring cells, are realized by the cooperation of intracellular and extracellular matters. For the molecules inside the cell, there are two main transporting forms: self-diffusion and directional transport. Self-diffusion is passive and depends on the cytoplasm; while directional transport is active, which can quickly realize material transport and signal conduction. It is worth mentioning that the receptor and ligand are an important class of material that transmits information, which regulates
basic tasks essential to life, such as  in cell growth, secretion, contraction, motility, and adhesion and  receptor-mediated cell behaviors \cite{Lauffenburger,Wiley,Lawley1,Lawley2,Lawley2020,Weikl,Abney}.

The biochemical receptors are proteins which bind the signal molecules and initiate responses in the target cell. The binding site of the receptor has a complex structure that is shaped to recognize the signal molecule with high specificity, helping to ensure that the receptor responds only to the appropriate signal and not to the many other signaling molecules surrounding the cell. In this case, the signal molecules are called ligands paired to the specific receptors. When the ligand binds to the receptor,  they become activated and generate various intracellular signals that alter the behavior of the cell \cite{Alberts}. As is known, receptor-ligand interactions play a major role in environmental sensing, signal transduction, and cell-to-cell signaling \cite{Zheng,Lines,Lin,Thewes}. Although receptors and ligands have many different forms, they come in closely matched pairs, with a receptor recognizing just one specific ligand, and a ligand binding to just one  target receptor.

 In many cases, receptors are transmembrane proteins on the target cell surface, while in other cases, the receptor proteins are inside the target cell which are either located in the cytoplasm and nucleus or in the membrane of subcellular compartments, such as the endoplasmic reticulum \cite{Lauffenburger}.  Receptors are uniquely able to direct  cell behavior by virtue of their ability to sense the environment, through binding of ligands, and their ability to transmit the signal to the cell interior, through interaction of their intracellular domains with enzymes and proteins within the cell. In this paper, we will focus on  the unbounded  receptors which bind to ligands on cell menbrane.

\begin{figure}[h!]
 \centering
\includegraphics[height=7cm ,width=10cm]{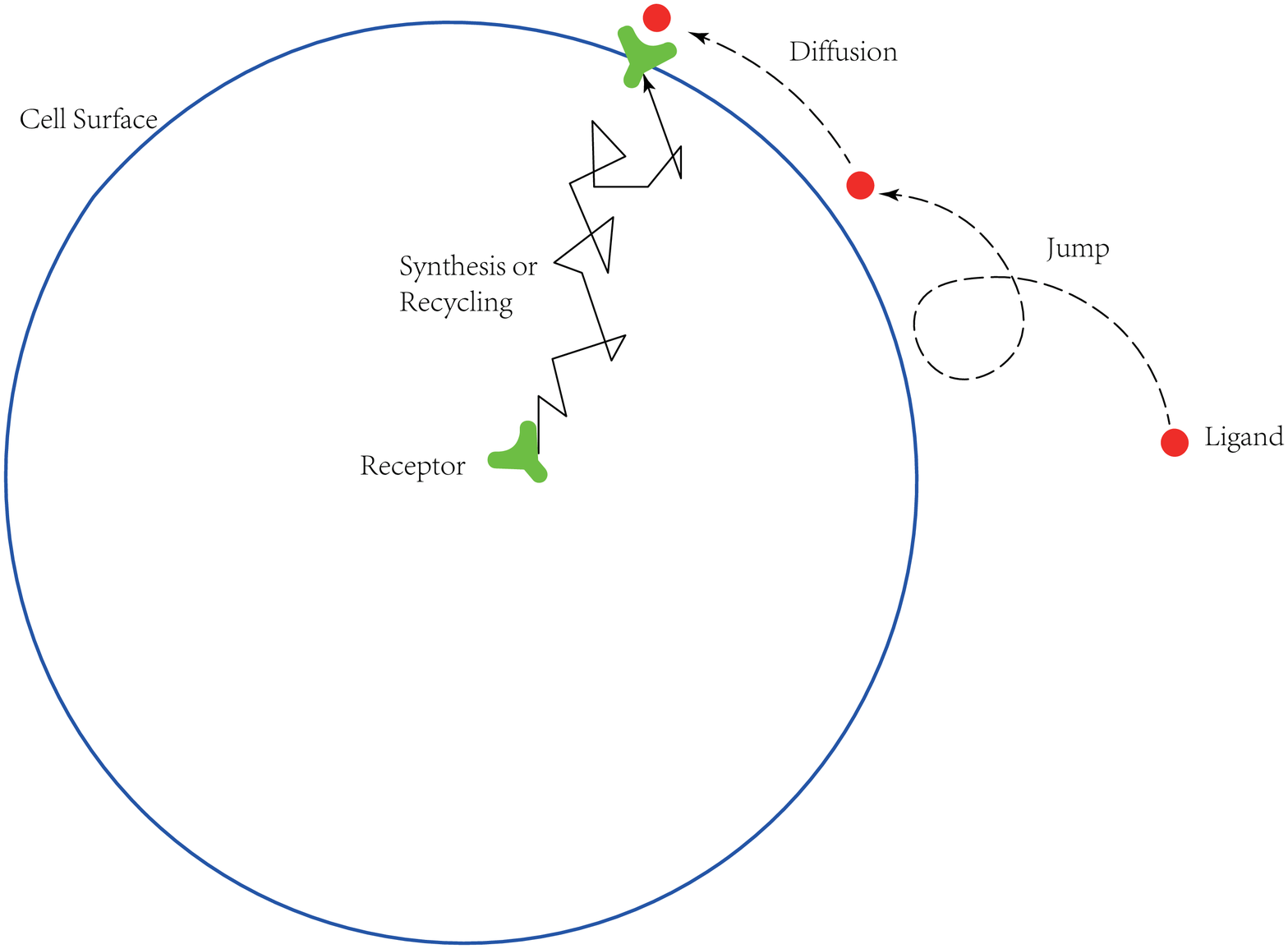}
 	\caption{(Color online) The illustration of receptor and ligand binding on cell membrane.}
 	\label{Fig.01}
 \end{figure}

 An additional influence on receptor-ligand binding on the cell surface is receptor mobility, or diffusion within the cell. After the recognition of a receptor on cell surface, the receptor-ligand complex will go into the endocytic cycle in the cell \cite{Pastan}, which governs the recycling to the cell surface (receptors) and delivery to lysosomes (ligands). Besides, the receptors are protein molecules which synthesize intracellularly and transport to the cell surface. It is the combined work that makes the change of receptors numbers on the cell surface, which undoubtedly affects the cellular responses to outside signals. As the intracellular transport processes can alter the number of receptors present on the cell surface, it is necessary to take the motion of receptors into consideration when examining the binding dynamics.
 The intracellular receptors are a class of receptors that locate in the cytoplasm and nucleus or in the membrane of subcellular compartments. They can pick up signals from outside the cell, bind to ligands that can be paired, trigger reactions inside the cell, and produce specific effects \cite{Bressloff2014,Bialek}. Diffusion effects  play a role in ligand binding to receptors \cite{Dobramysl,Linn}.
 Here we consider a reduced model by treating a cell as a disk, and the illustration can be seen in Figure \ref{Fig.01}.

 Dynamical systems arising in biophysics are often subject to random fluctuations. The noisy fluctuations may be Gaussian or non-Gaussian, which are modeled by Brownian motion or $\alpha-$stable L\'evy motion, respectively. There are various studies about Gaussian noise in receptors and ligands \cite{Basnayake}.
 However, as the ligand search for the surface receptor in the form of jump-diffusion \cite{Vauquelin}, which can not be modeled by purely diffusion any more, using the L\'evy motion with non-Gaussian distribution to depict its motion will be more appropriate. Indeed? many researchers have begun to consider non-Gaussian noise in cell biology \cite{Bressloff2014, Lomholt2008}. We will study the dynamics of ligands under non-Gaussian fluctuations. It is worth mentioning that noise-induced transition phenomena were regarded in various dynamical systems, including stochastic gene expression \cite{Wanghui2019,Hu}, climatology of the earth \cite{zhengyayun}, chemical reacting systems \cite{dy}. In particular, the most probable transition pathway attracts a lot of attention \cite{Hu,Duan,Li,Lu,Cai}.

 There are several approaches  for distinguishing the receptor and ligand binding events, including performing experiments at lowered temperatures and utilizing   pharmacologic agents. However, the existing studies mostly consider the binding of ligand and receptor when the receptor is confined to the cell membrane \cite{Vauquelin,Hu2015},   the dynamical behaviors of the receptor and ligand themselves are not considered. In this paper, we apply tools from stochastic dynamical systems to investigate receptor-ligand binding, where receptors and ligands are randomly moving.  We will examine the possibility of the receptor and ligand binding on the membrane, by calculating the joint probability of the receptor occurring on the membrane and ligand locating at the same site. Simultaneously, as  a receptor can appear at any location on the membrane, we will focus on the most probable trajectory of a receptor and find its probability arrival at the membrane.
 For the ligand, we examine the probability of the ligand reaching the domain close to the receptor on the cell membrane by means of Fokker-Planck equation.
 Then,  we relate the probability of the receptor diffusing to the cell membrane with the probability of the ligand arriving at the cell membrane to calculate the most probable binding probability.

 This paper is organized as follows. In Section 2, we present the mathematical model and compute the probability of the receptor diffusing to the cell membrane and the probability of the ligand arriving at the cell membrane respectively. In Section 3, we calculate the most probable binding probability. Finally, we end with some discussions in Section 4.

\section{Binding of a ligand to a receptor on cell membrane }
The binding event can be regarded as the encounter of a ligand and a receptor on the cell membrane. To investigate the binding dynamics, we consider the intracellular motion of the receptor and the extracellular motion of the ligand respectively. As the receptors are large protein molecules diffusing in cell in the form of vesicles and the intracellular environment is crowded, the motion of receptors can be treated as Brownian motion on a certain time scale  \cite{Bouchaud, Saffman, Havlin}.

\subsection{The probability of a cell-surface receptor diffusing to the cell membrane}

The intracellular motion of the receptor can be described by a Langevin equation with Brownian motion \cite{Bressloff2014}. By assuming the cell to be a 2-dimensional disk, we can rewrite the motion equation of receptor in the form:
\begin{equation}\label{eq.(2)}
	\begin{cases}
		dx=b_{1}(x,y)dt+\sqrt{2D_{1}}dB_{1}(t), \\
		dy=b_{2}(x,y)dt+\sqrt{2D_{2}}dB_{2}(t),
	\end{cases}
\end{equation}
where $b_1(x,y) ,b_2(x,y) $ represent the external force, $D_1$ and$ D_2$ are diffusion coefficients with respect to the Brownian motion, and $B_{t}=(B_{1}(t), B_{2}(t))$. Transforming Eq.(\ref{eq.(2)}) from Cartesian coordinates to polar coordinates by setting $x=r\cos\theta$ and $y=r \sin\theta$, we have

\begin{equation}
	\begin{cases}
		dx=-r \sin\theta d\theta+\cos\theta dr,\\
		dy=r \cos\theta d\theta+\sin\theta dr.\nonumber
	\end{cases}
\end{equation}
Thus Eq.(\ref{eq.(2)}) can be rewritten as
\begin{equation}\label{eq.(3)}
	{ \left(
		\begin{array}{cc}
			\cos\theta  & -r \sin\theta  \\
			\sin \theta   & r \cos\theta
		\end{array}
		\right)}{
		\left(
		\begin{array}{c}
		dr \\
			d\theta
		\end{array}
		\right)}={
		\left(
		\begin{array}{c}
			b_{1}(x,y)dt\\
			b_{2}(x,y)dt
		\end{array}
		\right)}+{
		\left(
		\begin{array}{cc}
			\sqrt{2D_{1}} & 0 \\
			0 & \sqrt{2D_{2}}
		\end{array}
		\right)}{
		\left(
		\begin{array}{c}
	dB_{1}(t)\\
			dB_{2}(t)
		\end{array}
		\right),}
\end{equation}
which leads to an evolution equation about $r$ and $\theta$ read as
\begin{equation}\label{eq.(4)}
	\left(
		\begin{array}{c}
			dr\\
			d\theta
		\end{array}
		\right)=B
		\left(
		\begin{array}{c}
			b_{1}(x,y)dt\\
			b_{2}(x,y)dt
		\end{array}
		\right)
	+B
		\left(
		\begin{array}{cc}
			\sqrt{2D_{1}} & 0 \\
			0 & \sqrt{2D_{2}}
		\end{array}
		\right)
		\left(
		\begin{array}{c}
			dB_{1}(t)\\
			dB_{2}(t)
		\end{array}
		\right),
\end{equation}
with $B$ denoting the inverse of the coefficient matrix in the left hand sides of Eq.(\ref{eq.(3)}), which has forms
\begin{equation}
	B= {
		\left(
		\begin{array}{cc}
			\cos\theta  & \sin\theta  \\
			-\sin\theta/r & \cos\theta/r \nonumber
		\end{array}
		\right ).
	}
\end{equation}
Eq.(\ref{eq.(4)}) gives us the Langevin equation in the polar coordinates.  For simplicity, we further assume the intracellular force $f$ in polar coordinates to be $\mathbf{r}=(r,\theta)\triangleq f$ \cite{Holcman}. The explicit choice of $f$ above is only to simplify many technicalities; the method used will apply to more general but intracellular force. So the Langevin equation Eq.(\ref{eq.(4)}) can be reduced as
\begin{equation}\label{eq.(5)}
	\left\{
	\begin{aligned}
			& dr= rdt+\sqrt{2D_{1}}\cos\theta dB_{1}(t)+\sqrt{2D_{2}}\sin\theta dB_{2}(t),  \\
			& d\theta=\theta dt-\sqrt{2D_{1}}\sin\theta/r dB_{1}(t)+\sqrt{2D_{2}}\cos\theta/r dB_{2}(t).
	\end{aligned}
	\right.
\end{equation}

In the practical biological processes, the receptor can locate many possible sites on the cell membrane. This indicates there could exist a mass of pathways for the receptor transiting to these sites. We are interested in the most probable transition pathways among all the possible trajectories. Based on Eq.(\ref{eq.(5)}), we will get the most probable pathway for a receptor starting from the center$(r,\theta)=(0,0)$ to the membrane $(r,\theta)=(1, \hat{\theta})$.
The probability of receptor occurring at site $\theta$ on membrane can be obtained by the Onsager-Machlup (OM) action functional \cite{Fujita,Capitaine}

\begin{equation}\label{eq.(6)}
	S(z,\dot{z})=\frac{1}{2}\int_{0}^{T}[\dot{z}-f(z)][\sigma\cdot\sigma^{T}(z)]^{-1}[\dot{z}-f(z)]+\bigtriangledown_{x}\cdot f(z) dt,
\end{equation}
with the Lagrangian
\begin{equation}\label{eq.(7)}
	\begin{split}
		L(z,\dot{z})&=\frac{1}{2}[\dot{z}-f(z)][\sigma\cdot\sigma^{T}(z)]^{-1}[\dot{z}-f(z)]+\frac{1}{2}\bigtriangledown_{x}\cdot f(z)\\
		&=\frac{1}{2}|D^{-1}[\dot{z}-f(z)]|^{2}+\frac{1}{2}\bigtriangledown_{x}\cdot f(z),
	\end{split}
\end{equation}
where $D$ denoting the matrix in Eq.(\ref{eq.(4)})
\begin{equation}
	D= B
		\left(
		\begin{array}{cc}
			\sqrt{2D_{1}} & 0 \\
			0 & \sqrt{2D_{2}}
		\end{array}
		\right)
	=
		\left(
		\begin{array}{cc}
			\sqrt{2D_{1}}\cos\theta  & \sqrt{2D_{2}}\sin\theta  \\
				-\sqrt{2D_{1}}\sin\theta/r & \sqrt{2D_{2}}\cos\theta/r    \nonumber
				\end{array}
				\right ).
		\end{equation}
Thus the Lagrangian becomes \cite{chaoying}
\begin{equation}\label{eq.(8)}
	L(z,\dot{z})=\frac{1}{2}(\frac{\cos^{2}\theta}{2D_{1}}+\frac{\sin^{2}\theta}{2D_{2}})(r-\dot{r})^{2}+\frac{1}{2}(\frac{r^{2}\sin^{2}\theta}{2D_{1}}+\frac{r^{2}cos^{2}\theta}{2D_{2}})(\theta-\dot{\theta})^{2}+1.
\end{equation}
According to the Euler-Lagrange equation
\begin{equation}
	\frac{d}{dt}\frac{\partial}{\partial\dot{z}}L(z,\dot{z})=\frac{\partial}{\partial z}L(z,\dot{z}),
\end{equation}
we get the following second order differential equations for $r$ and $\theta$

\begin{equation}\label{eq.(10)}
	\left\{
	\begin{aligned}
	&\ddot{r}=r+r\theta^{2}(\frac{\sin^{2}\theta}{D_{1}}+\frac{\cos^{2}\theta}{D_{2}})/(\frac{\cos^{2}\theta}{D_{1}}+\frac{\sin^{2}\theta}{D_{2}}),   \\
	&\ddot{\theta}=\theta+\sin\theta\cos\theta(\frac{1}{D_{2}}-\frac{1}{D_{1}})(1-\theta^{2})/(\frac{\sin^{2}\theta}{D_{1}}+\frac{\cos^{2}\theta}{D_{2}}).
	\end{aligned}
	\right.
\end{equation}
This is rewritten as the following four dimensional system
\begin{equation}
	\label{eq.(11)}
	\left\{
	\begin{aligned}
		& \dot{r}=w, \\
		& \dot{\theta}=v,  \\
		&  \dot{w}=r+r\theta^{2}(\frac{\sin^{2}\theta}{D_{1}}+\frac{\cos^{2}\theta}{D_{2}})/(\frac{\cos^{2}\theta}{D_{1}}+\frac{\sin^{2}\theta}{D_{2}}),   \\
		&  \dot{v}=\theta+\sin\theta \cos\theta(\frac{1}{D_{2}}-\frac{1}{D_{1}})(1-\theta^{2})/(\frac{\sin^{2}\theta}{D_{1}}+\frac{\cos^{2}\theta}{D_{2}}).
	\end{aligned}
	\right.
\end{equation}

As we aim at getting the most probable transition pathway in a fixed period of time and its corresponding Onsager-Machlup functional action value $S(z,\dot{z})$, as well as the probability, the next step will be reduced to solve a two-point boundary problem Eq.(\ref{eq.(11)}) with initial location  $(r(0),\theta(0))=(0,0)$ and final location $(r(T),\theta(T))=(1,\hat{\theta})$ where $ \hat{\theta}\in (0,2\pi)$. Here $T$ is the transition time for a receptor from the cytoplasm to cell membrane which we consider as a fixed period of time and can be set $T=5$.

Assuming that the receptor starting inside the cell can reach any site on the membrane, we divide the membrane (the circle) into 8 equal segments: $(0,\frac{\pi}{4}),(\frac{\pi}{4},\frac{\pi}{2}),...$, and find the most probable transition pathway on each segment as well as the minimal action value. The key lies in solving  Eq.(\ref{eq.(11)}), but the only known conditions are the boundary value so for with the initial velocity still unknown. So here we adopt a neural network approach to handle the tricky problem. The main idea is to construct a neural network and approximate the mapping from the final point to the initial velocity, such that the solution of Eq.(\ref{eq.(11)}) with this initial velocity at time $T$ meets the target given final point. This method has been used in solving differential equations with boundary problems which is called neural shooting \cite{ Ibraheem}. Generally, the algorithm can be summarized into the following steps:

\noindent
\textbf{Step 1:} Generate the data sets to supply the training basis for this  neural network, by simulating the Eq.(\ref{eq.(11)}) given a series of initial velocities with setting the noisy intensity parameters $D_1=D_2=0.01$. Then we get a set of final points at time $T$  which corresponds to a set of initial velocities. This dataset consists of final point-velocity pairs.

\noindent
\textbf{Step 2:} Adopt the neural network ``2-8-16-8-2" constructed by Tensorflow to approximate the mapping from the initial velocity to the final point at time $T$ by using the dataset generated in Step 1.

\noindent
\textbf{Step 3:} Use the learned model in Step 2 to predict the initial velocities given the possible final points on the surface, and insert the initial velocity into the system to compute the transition pathway as well as its corresponding Onsager-Machlup action functional value.

The Onsager-Machlup action functional value of each pathway on the divided segments is calculated in Table 1-8 and the minimum action functional value is selected with bold font, that is the most probable transition pathway we are looking for.

\begin{table}[!htb]
	\centering
	\small
	\caption{\ The action functional value of pathways terminated on $(0,\pi/4)$ with predicted initial velocity $(w_{0},v_{0})$}.
	\label{tb1}
	\begin{tabular*}{1.0\textwidth}{@{\extracolsep{\fill}}lllllllllll}
		\hline
		$\hat{\theta}$ & 0.1 &  0.2 &  \textbf{0.3} & 0.4 & 0.5 & 0.6 & 0.7 & $\pi/4$  \\
        \hline 
        $w_{0}$ & 0.1030 & 0.1024 & 0.1016 & 0.1005 & 0.0991 & 0.0974 & 0.0956 &0.0935
        \\
        \hline
        $v_{0}$ & 0.0096 &  0.0196 & 0.0297 & 0.0398 & 0.0498 &  0.0599 & 0.0700 & 0.0801
        \\
 	\hline
		$S$ & 3.99 & 3.92 & \textbf{3.88} & 3.89 & 3.97 & 4.16 & 4.65 & 7.12\\
		\hline
	\end{tabular*}
\end{table}

\begin{table}[!htb]
	\centering
	\small
	\caption{\ The action functional value of pathways terminated on $(\pi/4,\pi/2)$  with predicted initial velocity $(w_{0},v_{0})$}.
	\label{tb2}
	\begin{tabular*}{1.0\textwidth}{@{\extracolsep{\fill}}lllllllllll}
		\hline
		$\hat{\theta}$ & \textbf{0.9} &  1.0 &  1.1 & 1.2 & 1.3 & 1.4 & 1.5 & $\pi/2$  \\
        \hline 
        $w_{0}$ & 0.0882 & 0.0855 & 0.0827 & 0.0797 & 0.0766 & 0.0733 & 0.0700 &0.0666
        \\
        \hline
        $v_{0}$ & 0.0934 &  0.1034 & 0.1134 & 0.1234 & 0.1334 & 0.1434 & 0.1534 & 0.1634
        \\
		\hline
		$S$ & \textbf{28.83} & 348.86 & 1568.5 & 498.76 & 99.31 & 63.05 & 216.42 & 123.02\\
		\hline
	\end{tabular*}
\end{table}

\begin{table}[!htb]
	\centering
	\small
	\caption{\ The action functional value of pathways terminated on $(\pi/2,3\pi/4)$
 with predicted initial velocity $(w_{0},v_{0})$.}
	\label{tb3}
	\begin{tabular*}{0.9\textwidth}{@{\extracolsep{\fill}}lllllllllll}
		\hline
		$\hat{\theta}$ & 1.7 &  1.8 &  1.9 & 2.0 & 2.1 &\textbf{ 2.2} & 2.3 & $3\pi/4$  \\
       \hline 
        $w_{0}$ & 0.0617 & 0.0582 & 0.0548 & 0.0513 & 0.0478 & 0.0444 & 0.0411 &0.0378
        \\
        \hline
        $v_{0}$ & 0.1711 &  0.1814 & 0.1917 & 0.2020 & 0.2123 & 0.2226 & 0.2329 & 0.2432
        \\
		\hline
		$S$ & 41.86 & 28.54 & 373.05 & 17.34 & 435.13 & \textbf{16.48} & 177.57 & 305.81\\
		\hline
	\end{tabular*}
\end{table}

\begin{table}[!htb]
	\centering
	\small
	\caption{\ The action functional value of pathways terminated on $(3\pi/4,\pi)$}
	\label{tb4}
	\begin{tabular*}{0.9\textwidth}{@{\extracolsep{\fill}}lllllllllll}
		\hline
		$\hat{\theta}$ & 2.4 &  2.5 &  2.6 & 2.7 & 2.8 & \textbf{2.9} & 3.0 & $\pi$  \\
		\hline
    $w_{0}$ &0.0454 & 0.0423 &0.0393  &0.0363  &0.0333  & 0.0306 & 0.0279 &0.0253
        \\
        \hline
        $v_{0}$ & 0.2408 &  0.2509 & 0.2611 & 0.2711 & 0.2811 & 0.291 & 0.301 & 0.3108
        \\
		\hline
		$S$ & 197.84 & 320.53 & 208.36 & 804.55 & 1055.6 & \textbf{161.68} & 224.56 & 1906.5\\
		\hline
	\end{tabular*}
\end{table}

\begin{table}[!htb]
	\centering
	\small
	\caption{\ The action functional value of pathways terminated on $(-\pi/4,0)$}
	\label{tb5}
	\begin{tabular*}{0.9\textwidth}{@{\extracolsep{\fill}}lllllllllll}
		\hline
		$\hat{\theta}$ & $-\pi/4$ & -0.6 &  -0.5 & -0.4 & -0.3 & -0.2 & \textbf{-0.1}  \\
		\hline
  $w_{0}$ &0.0957 & 0.0974 &0.0989 &0.1002  &0.1012 & 0.102 & 0.1024
        \\
        \hline
        $v_{0}$ & -0.0692 &  -0.0593 & -0.0494 & -0.0396 & -0.0297 & -0.0197 & -0.0099
        \\
		\hline
		$S$ & 12.75 & 7.96 & 6.22 & 5.36 & 4.84 & 4.52 & \textbf{4.29}\\
		\hline
	\end{tabular*}
\end{table}
\begin{table}[!htb]
	\centering
	\small
	\caption{\ The action functional value of pathways terminated on $(-\pi/2,-\pi/4)$}
	\label{tb6}
	\begin{tabular*}{0.9\textwidth}{@{\extracolsep{\fill}}lllllllllll}
		\hline
		$\hat{\theta}$ & $-\pi/2$ &  -1.5 & \textbf{-1.4} & -1.3 & -1.2 & -1.1 & -1.0 & -0.9 \\
		\hline
  $w_{0}$ &0.0747 & 0.0773 &0.0799 &0.0824  &0.0849  & 0.0872 & 0.0896 &0.0917
        \\
        \hline
        $v_{0}$ & -0.1595 &  -0.1494 & -0.1393 &-0.1293 & -0.1192 & -0.1092 & -0.0992 & -0.0892
        \\
		\hline
		$S$ & 5193.3 & 235.1 & \textbf{204.79} & 996.73 & 1114.8 & 774.8 & 435.49 & 7503.1\\
		\hline
	\end{tabular*}
\end{table}
\begin{table}[!htb]
	\centering
	\small
	\caption{\ The action functional value of pathways terminated on $(-3\pi/4,-\pi/2)$}
	\label{tb7}
	\begin{tabular*}{0.9\textwidth}{@{\extracolsep{\fill}}lllllllllll}
		\hline
		$\hat{\theta}$ & $-3\pi/4$ &  -2.3 &  \textbf{-2.2} & -2.1 & -2.0 & -1.9& -1.8 & -1.7 \\
		\hline
    $w_{0}$ &0.0522 & 0.0547 &0.0573 &0.0599  &0.0625  & 0.0652 & 0.0679 &0.0706
        \\
        \hline
        $v_{0}$ & -0.2314 &  -0.2312 & -0.221 &-0.2107 & -0.2005 & -0.1904 & -0.1802 & -0.17
        \\
		\hline
		$S$ & 104.15 & 7134.9 & \textbf{50.41} & 4051.5 & 68.02 & 387.07 & 148.92 & 131.47\\
		\hline
	\end{tabular*}
\end{table}

\begin{table}[!htb]
	\centering
	\small
	\caption{\ The action functional value of pathways terminated on $(-\pi,-3\pi/4)$}
	\label{tb8}
	\begin{tabular*}{0.9\textwidth}{@{\extracolsep{\fill}}lllllllllll}
		\hline
		$\hat{\theta}$ & $-\pi$ &  -3.1 &  -3.0 & \textbf{-2.9} & -2.8 & -2.7& -2.6 & -2.5 \\
		\hline
    $w_{0}$ &0.0342& 0.0547 &0.0384 &0.0406  &0.043  & 0.0455 & 0.048 &0.0507
        \\
        \hline
        $v_{0}$ & -0.3272 &  -0.317 & -0.3068 &-0.2966 & -0.2864 & -0.2761 & -0.2658 & -0.2556
        \\
		\hline
		$S$ & 76.65 & 103.4 & 522.58 & \textbf{58.17} & 294.69 & 452.05 & 66.9 & 633.63\\
		\hline
	\end{tabular*}
\end{table}

\begin{figure}[h!]
	\centering
	\includegraphics[height=9cm ,width=10cm]{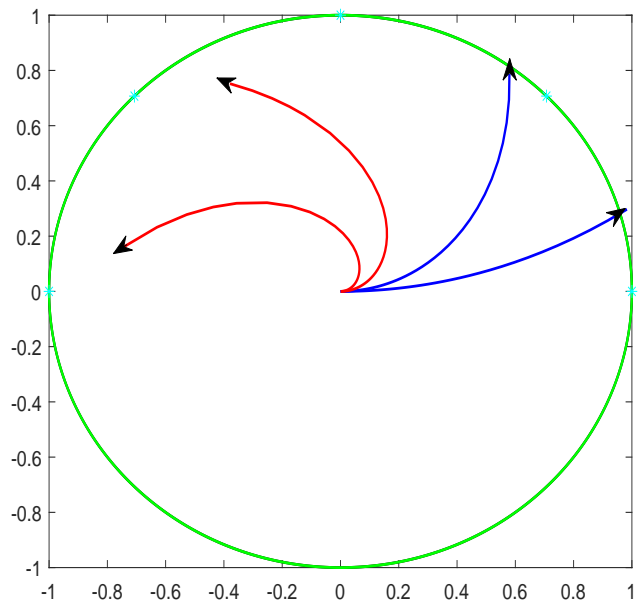}
	\caption{ The most probable transition pathways (with minimal action functional value) terminated on the upper half segments of the circle are shown in the figure. The blue pathways denote the reachable transitions, while the red ones fail to transit namely unreachable pathways. }
	\label{Fig.2}
\end{figure}

\begin{figure}[h!]
	\centering
	\includegraphics[height=9cm ,width=10cm]{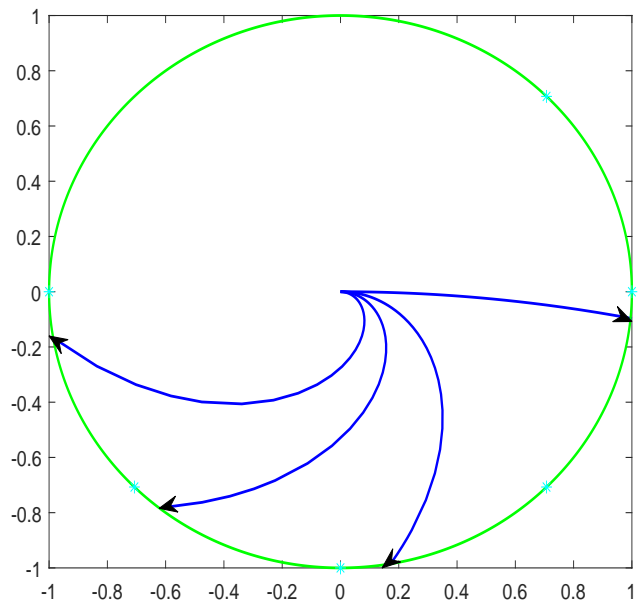}
	\caption{ The most probable  transition pathways (with minimal action  functional value) terminated on the half bottom of the circle. }
	\label{Fig.3}
\end{figure}

According to the minimal action functional value on each segment, we can get the  most probable transition pathways from the origin to the corresponding points on the circle. However, not all the pathways could reach exactly to the circle, because the pathways are obtained from the given initial velocities, which are estimated from the neural network algorithm. Even we have expanded the sample size and limited the predict error, there still exist pathways that did not satisfy the judgment condition whose final points were not exactly located on the circle. For these pathways, we call them ``unreachable". Tables 1-8 give us indication to find the most probable transition paths from the origin to the segments on surface and we show them in Figures \ref{Fig.2}-\ref{Fig.3}.

According to the Onsager-Machlup theory of stochastic dynamical systems, the probability of the transition pathway on a small tube can be estimated as
\begin{equation}
	P_{1}({||X-z||_{T}<\delta}|z(0)=(0,0),z(T)=(r,\theta))\approx c(T,\delta)e^{-S(z,\dot{z})}.
\end{equation}
When the parameters $\delta$ and $T$ are fixed, it can be seen that $P_{1}$ is only dependent on the action functional value, and the coefficient $c(T,\delta)$ can be regarded as a constant which does not vary with the final location on the membrane. As we are aiming at qualitatively exploring the difference of the probabilities transiting to each segment, here we get the approximated probability of the most probable transition pathway on each segment by ignoring the constant $c(T,\delta)$ in the following table:

\begin{table}[!htb]
	\centering
	\small
	\caption{\ The probability $P_{1}$ ($\propto e^{-S(z,\dot{z})}$) of most probable transition pathway on each segment}
	\label{tb9}
	\begin{tabular*}{1.0\textwidth}{@{\extracolsep{\fill}}lllllllllll}
		\hline
		$Segment$ &$ (0,\pi/4)$ &$ (\pi/4,\pi/2) $& $ (\pi/2,3\pi/4) $& $(3\pi/4,\pi)$  \\
		\hline
		$P_{1}$ &$ 0.02$ & $3.02\times 10^{-13}$ & $ 7.10\times 10^{-8}$ & $6.07\times 10^{-71}$ \\
		\hline\hline
		$Segment$ &$ (-\pi/4,0) $&$ (-\pi/2,-\pi/4)$ & $ (-3\pi/4,-\pi/2)$ &$ (-\pi,-3\pi/4) $ \\
		\hline
		$P_{1}$ &  $0.01$ & $1.15\times 10^{-89}$ & $1.28\times 10^{-22}$ & $ 5.46\times 10^{-26}$ \\
		\hline
	\end{tabular*}
\end{table}

It should be noted that the most probable transition pathways shown above are different from the real trajectories, as the calculated result is smooth curves obtained from solving the Euler-Lagrange equation, but it quantitatively points out the most possible reaching sites on the membrane, and there exist at least one trajectory around the most probable transition pathway.
\subsection{The probability of the ligand arriving at the cell membrane}
The intercellular motion of the ligand can be regarded as a jump-diffusion process along the cell membrane \cite{Vauquelin,Reynolds}. Ignoring the external forces of the external environment, which can be characterized by the L\'evy motion obeying $L^{\alpha}_{t}\sim S_{\alpha}(|t|^{1/\alpha},0,0)$ . The model can be represented as \cite{Duan}:

\begin{equation}
	dX_t=\sigma dB_t+\epsilon dL_t.
\end{equation}
The corresponding Fokker-Planck equation is
\begin{equation}
	\frac{\partial p(x,t) }{\partial t}=\sigma ^2 \Delta  p(x,t) +\epsilon^{\alpha} \frac{\partial ^{\alpha}}{\partial |x|^{\alpha}}p(x,t).
\end{equation}

The ligand locates on the site $\theta$ with probability
\begin{equation}
	P_{2}(\theta)\triangleq \int_{\theta-\delta}^{\theta+\delta} p(x,t)ds,
\end{equation}
here $P_{2}(\theta)$ can be calculated by integrating the density function $p(x,t)$ over$(\theta-\delta,\theta+\delta)$, $s$ is the arc length  between $(\theta-\delta)$ and $(\theta+\delta)$,  where $p(x,t)$ satisfies the Fokker-Planck equation. For computational convenience, we treat small segments on the cell membrane as one dimensional. In this case, there is no symmetry for the probability of the ligand arriving at the cell membrane. The fractional order derivation can be defined as follows \cite{Duan}:
\begin{equation}
	\begin{aligned}
		\frac{\partial ^{\alpha}}{\partial |x|^{\alpha}}p(x,t) &= -(-\Delta )^{\frac{\alpha}{2}}p(x,t) \\
		&=\int_{\mathbb{R}\backslash \{0\}}[p(x+y,t) - p(x,t) - I_{\{|y|<1\}}(y)yp'(x)]\nu_{\alpha}(dy).
	\end{aligned}
\end{equation}
This nonlocal equation can be numerically solved by a similar finite difference method as in \cite{Gao2016}.
We divide the cell membrane into the same parts as that in subsection 2.1 and list some computing results in Table \ref{tb10}, Table \ref{tb11}, Table \ref{tb12}. We also observe the evolution of  $P_{2}$ with related parameters as shown in the following figures.
\begin{figure}[h!]	
\subfigure[]{ \label{Fig.sub.31}
\includegraphics[width=0.5\textwidth]{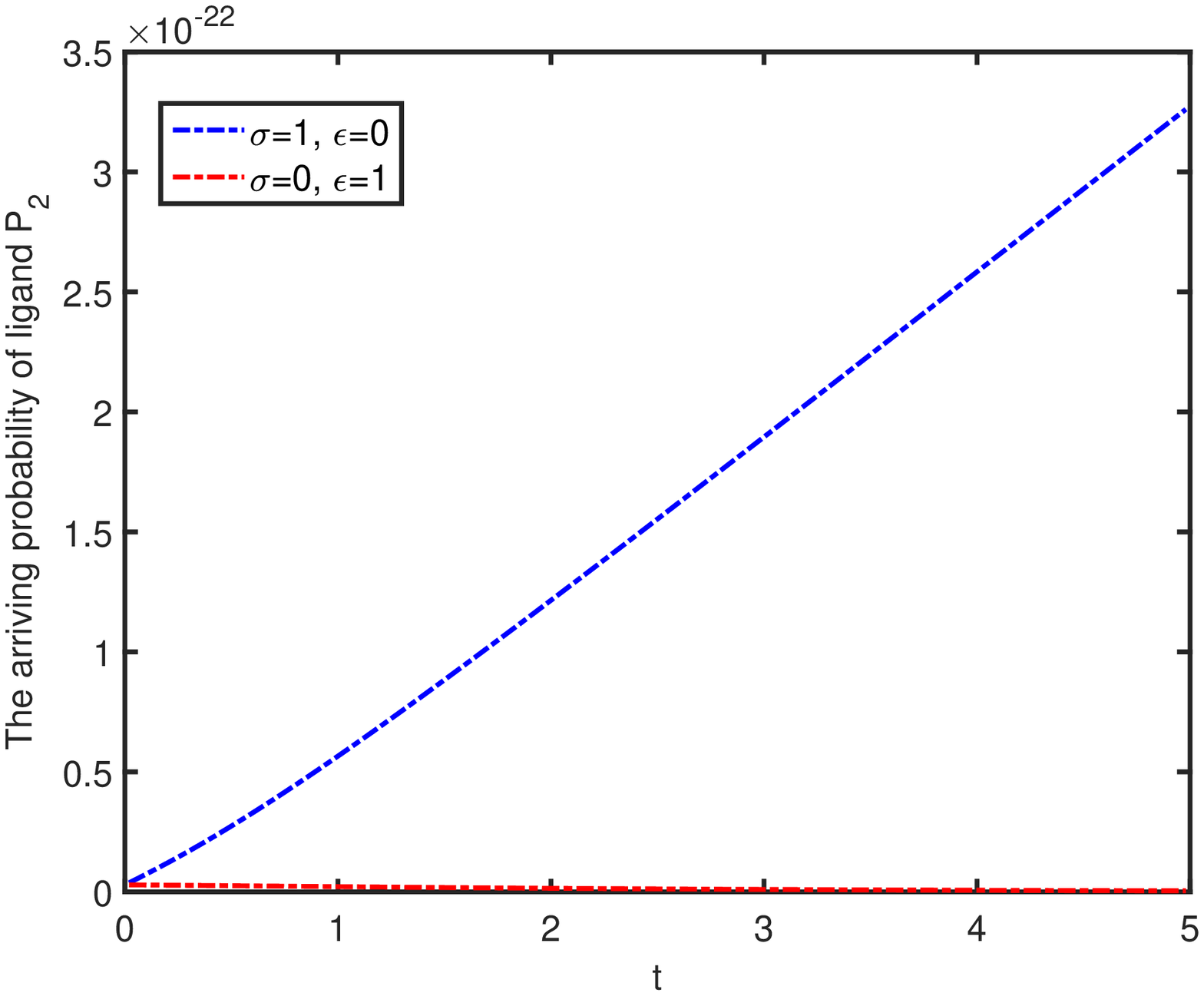}}
\subfigure[]{ \label{Fig.sub.32}
\includegraphics[width=0.5\textwidth]{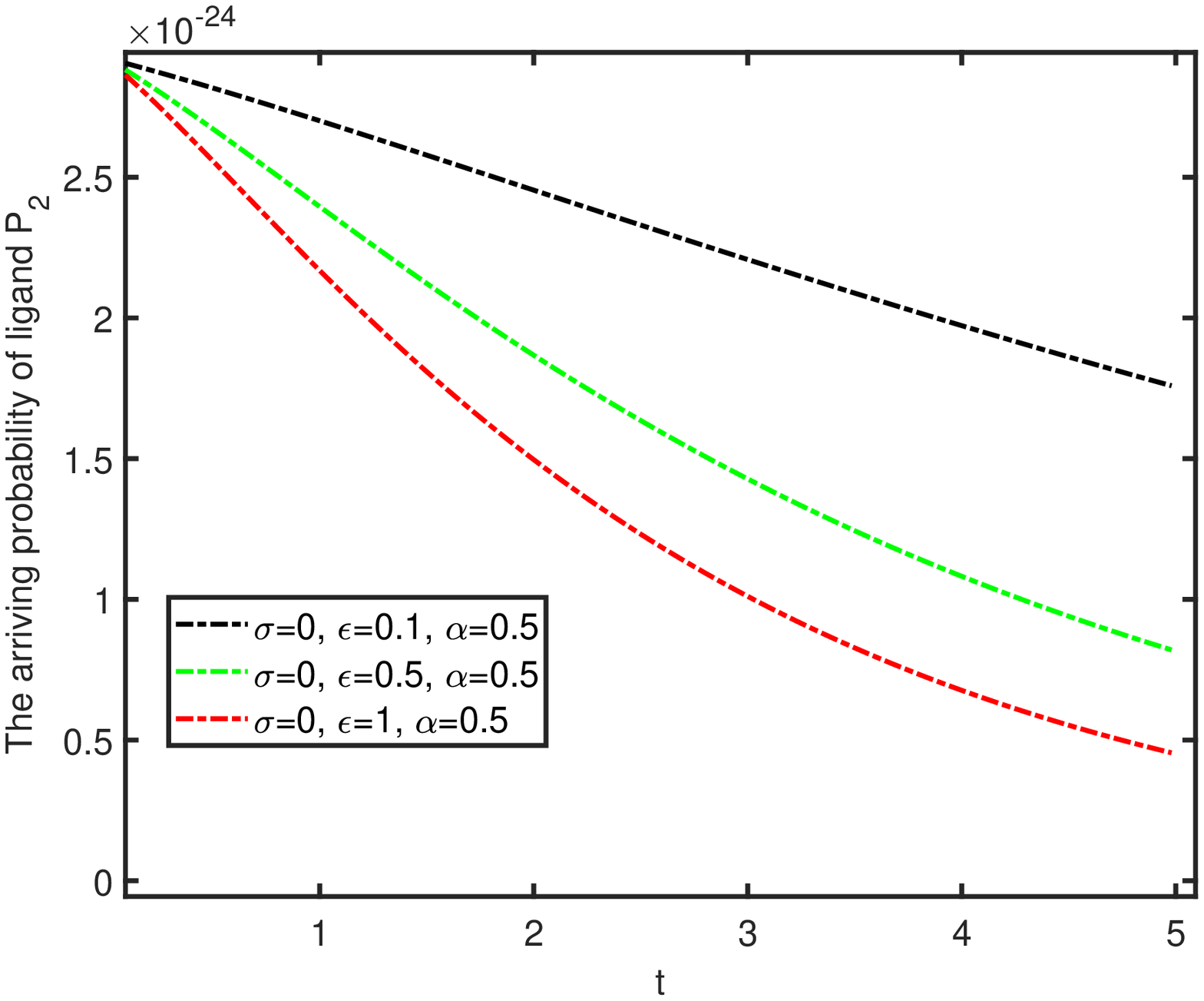}}
	\caption{ \textbf{(a) Compare Gaussian and non-Gaussian noise. (b) Evolution with different non-Gaussian noise intensity $\epsilon$: T=5, $\sigma=0$,  $\alpha=0.5$.}}
	\label{case01}
\end{figure}

In Figure \ref{case01}, we present the probability $P_{2}$ in the cases  of Brownian motion and L\'evy motion with different noise intensities. we draw it by setting $T=5$, $\sigma=1$  in $[0,\pi]$  domain. It shows that the probability $P_{2}$ with Gaussian noise is much bigger than that with non-Gaussian noise, which means that the ligand with purely diffusion is more likely to arrive at a given site on the surface than with jump-diffusion motion, and the larger the noise intensity of the L\'evy motion is, the less possibly the ligand can find the site. That can be explained by the difference between the two motions: the particles with Brownian motion find the binding site one by one and can exactly locate a site on the memnrane, while in the jump-diffusion case, the particles could jump over the site and wander around that, which will reduce the possibility of correct bindings.

\begin{figure}[h!]
	\subfigure[]{ \label{Fig.sub.01}
\includegraphics[width=0.5\textwidth]{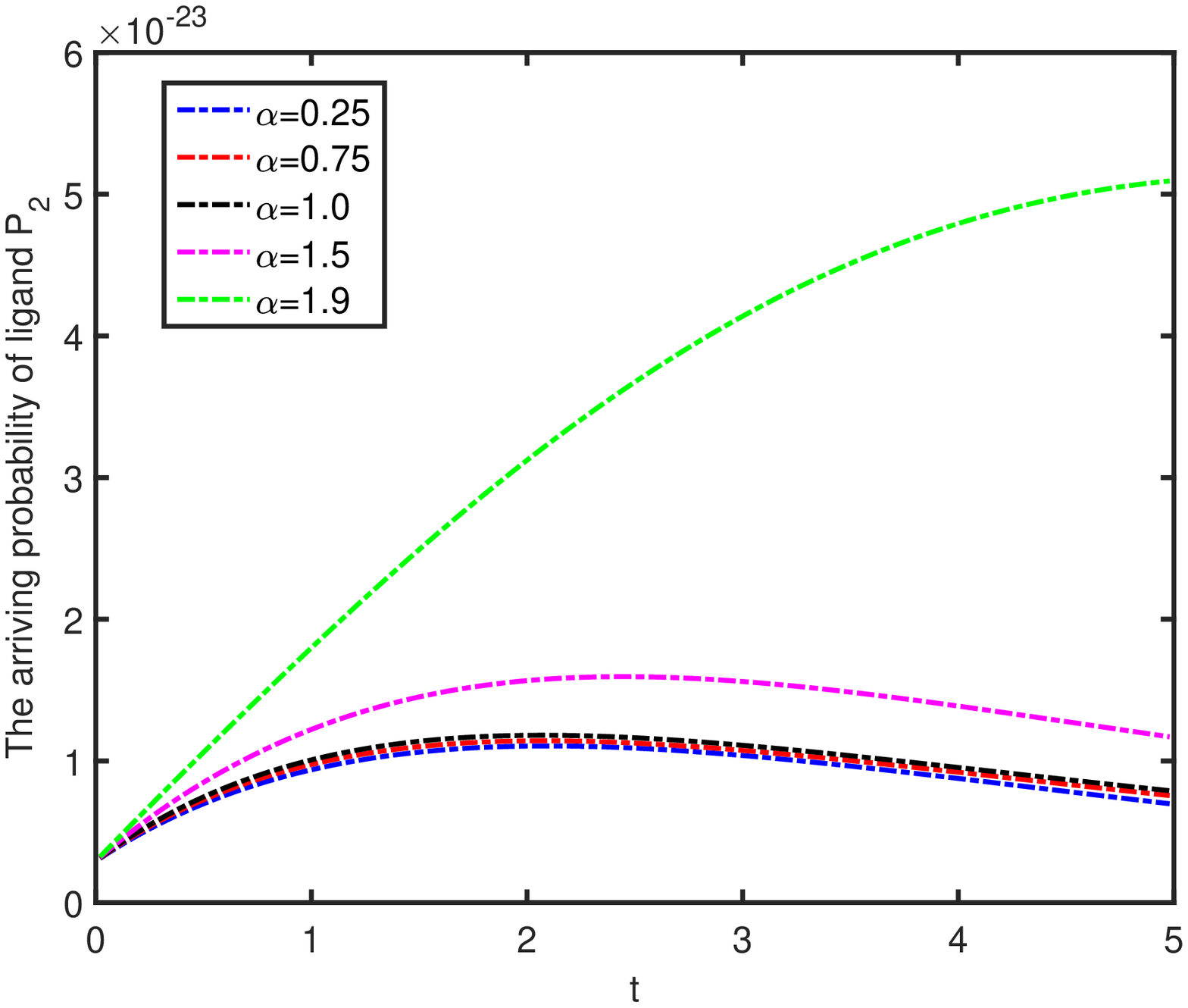}}
\subfigure[]{ \label{Fig.sub.02}
\includegraphics[width=0.5\textwidth]{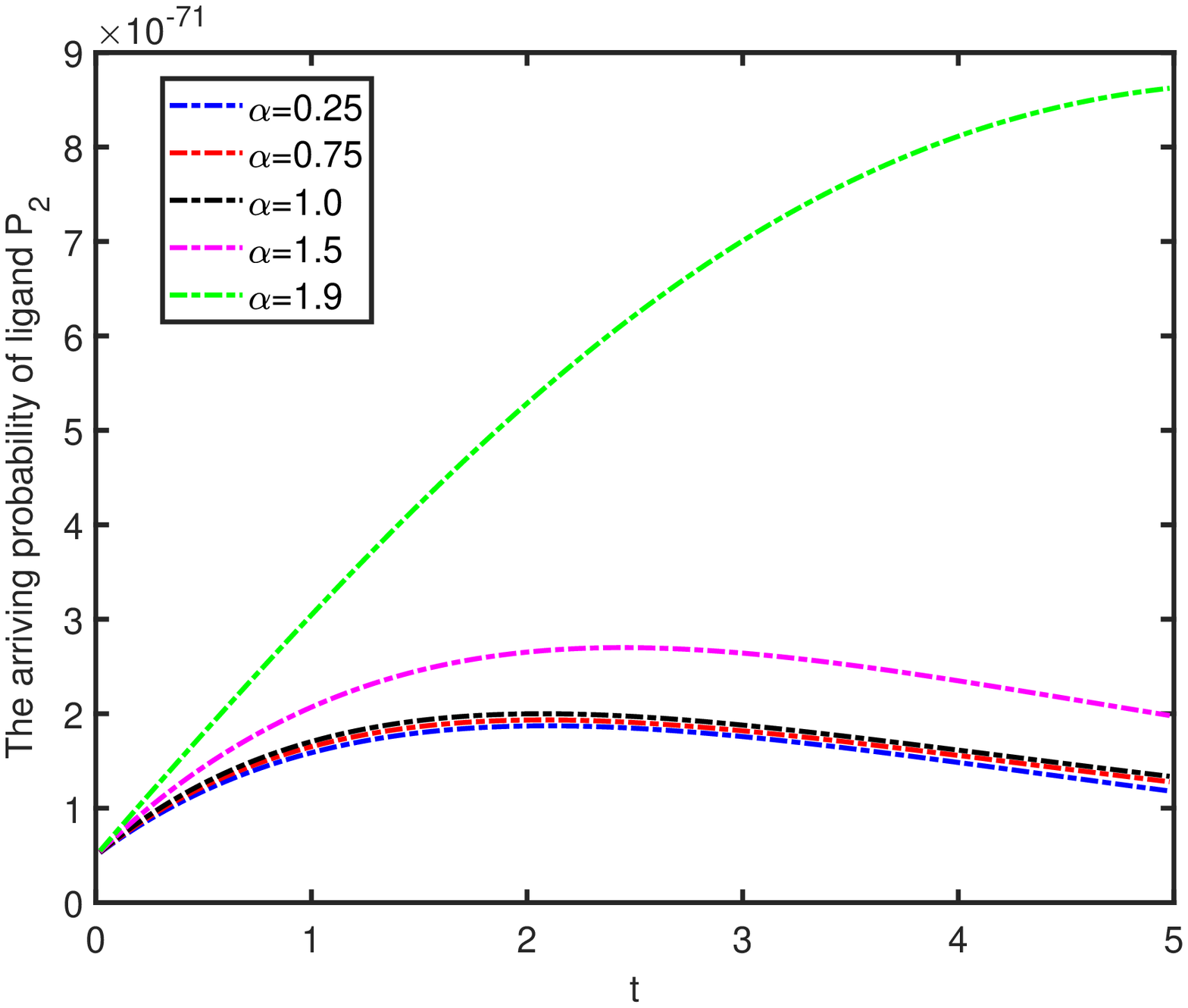}}
	\caption{\textbf{Evolution with different  non-Gaussian index of L\'evy process $\alpha$: T=5, $\sigma=0.25$,  $\epsilon=0.25$.} (a) $[0,\pi]$. (b)$[\pi,2\pi]$.}
	\label{case1}
\end{figure}

To explore the influence of non-Gaussian index on probability $P_{2}$, we paint Figure \ref{case1} by setting $T=5$, $\sigma=0.25$ and $\epsilon=0.25$ and consider that in $[0,\pi]$ and $[\pi,2\pi]$  domains separately. In  $[0,\pi]$, $P_{2}$  first increases  then decreases with respect to time $t$  and the non-Gaussian index of L\'evy process $\alpha$ has a significant effect on increasing the probability $P_{2}$, which also tells us that as time goes, it becomes difficult for the ligand to bind the given site, and the strong non-Gaussianity leads to a same result. In  $[0,\pi]$ and $[\pi,2\pi]$, $P_{2}$  has the similar results  in time.
\begin{figure}[h!]
	\subfigure[]{ \label{Fig.sub.1}
\includegraphics[width=0.5\textwidth]{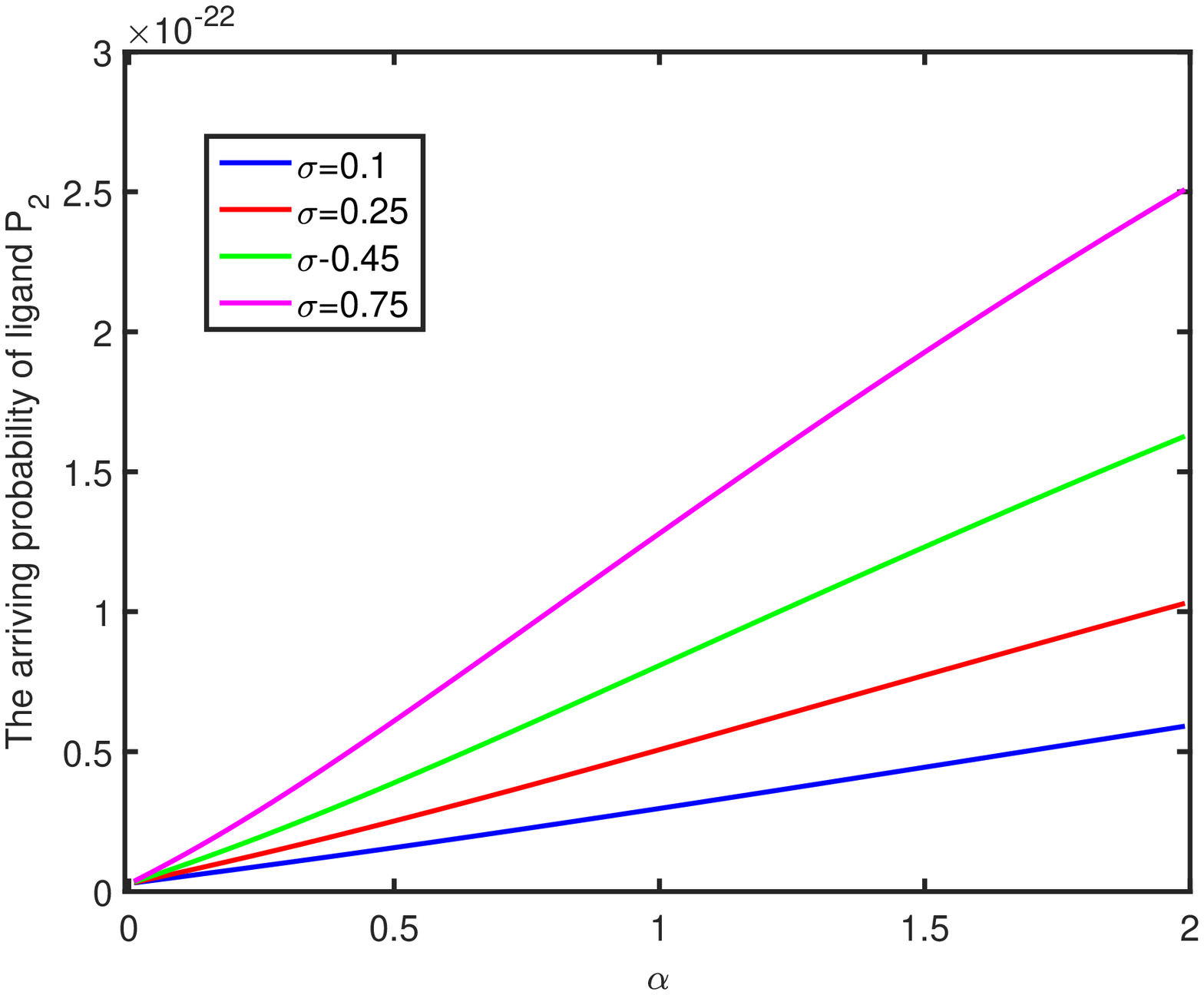}}
\subfigure[]{ \label{Fig.sub.2}
\includegraphics[width=0.5\textwidth]{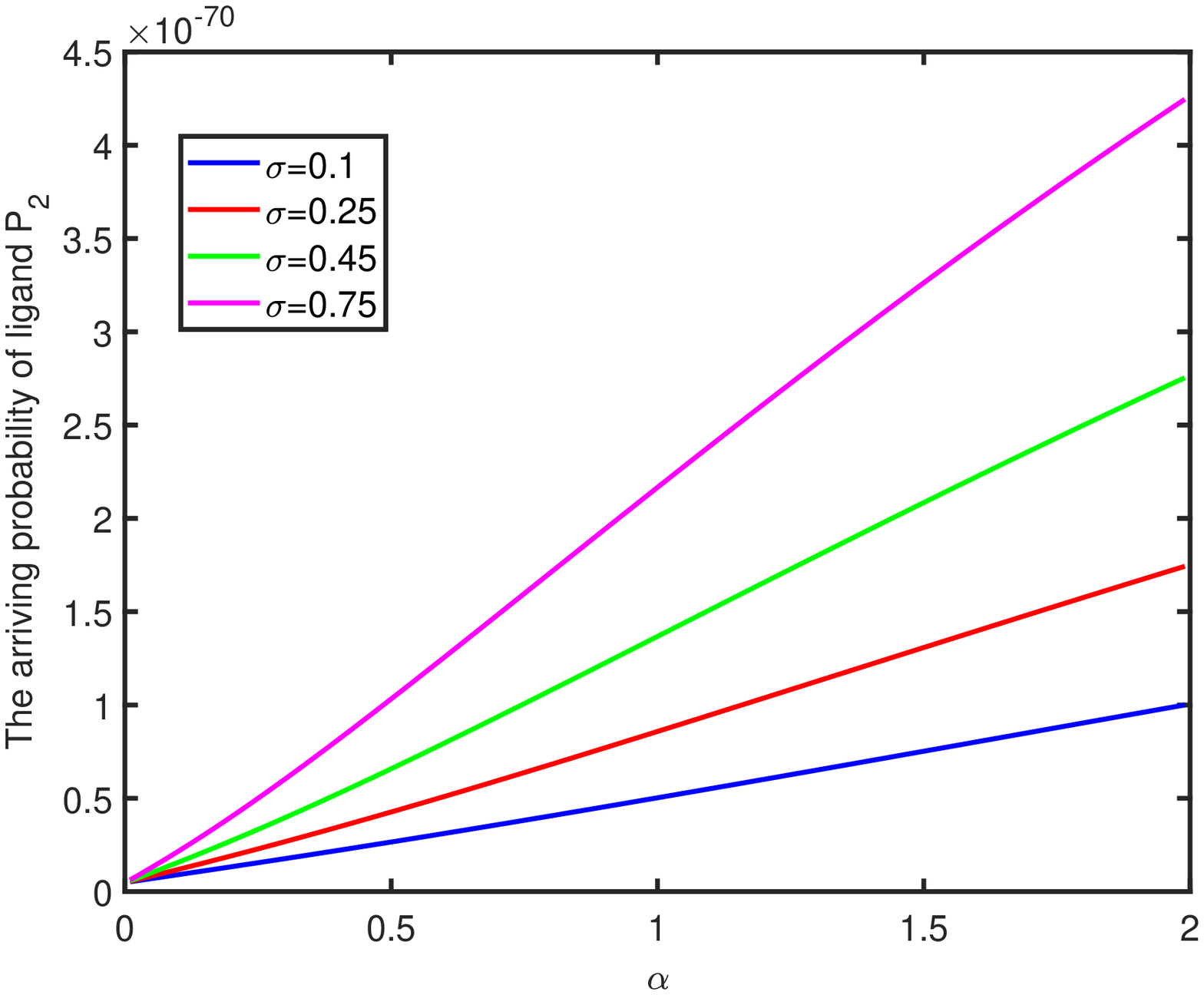}}
	\caption{\textbf{ Evolution with different Gaussian noise intensity $\sigma$: T=5, $\epsilon=0.25$.} (a) $[0,\pi]$. (b)$[\pi,2\pi]$.}
	\label{case2}
\end{figure}

\begin{figure}[h!]
\subfigure[]{ \label{Fig.sub.11}
\includegraphics[width=0.5\textwidth]{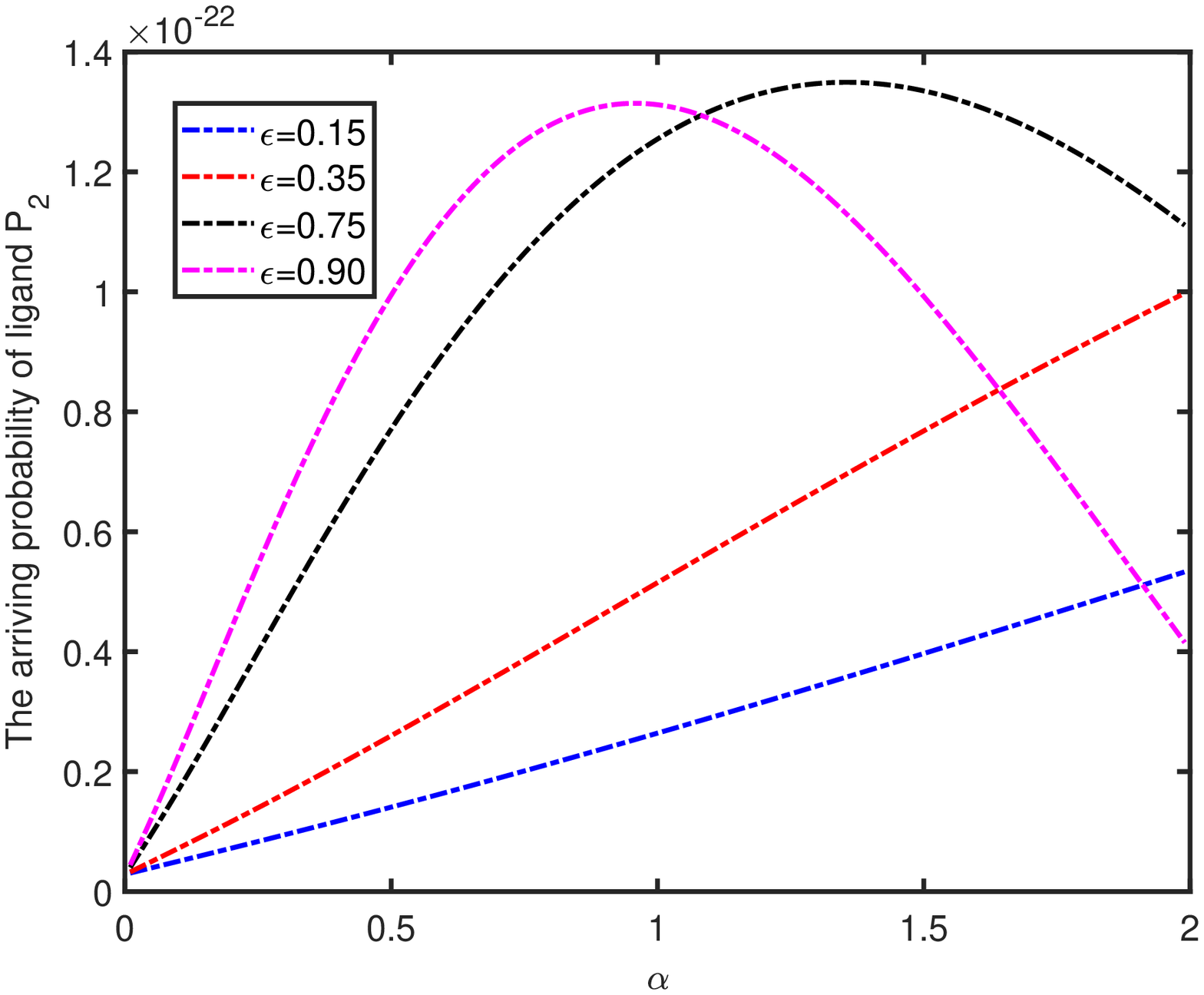}}
\subfigure[]{ \label{Fig.sub.12}
\includegraphics[width=0.5\textwidth]{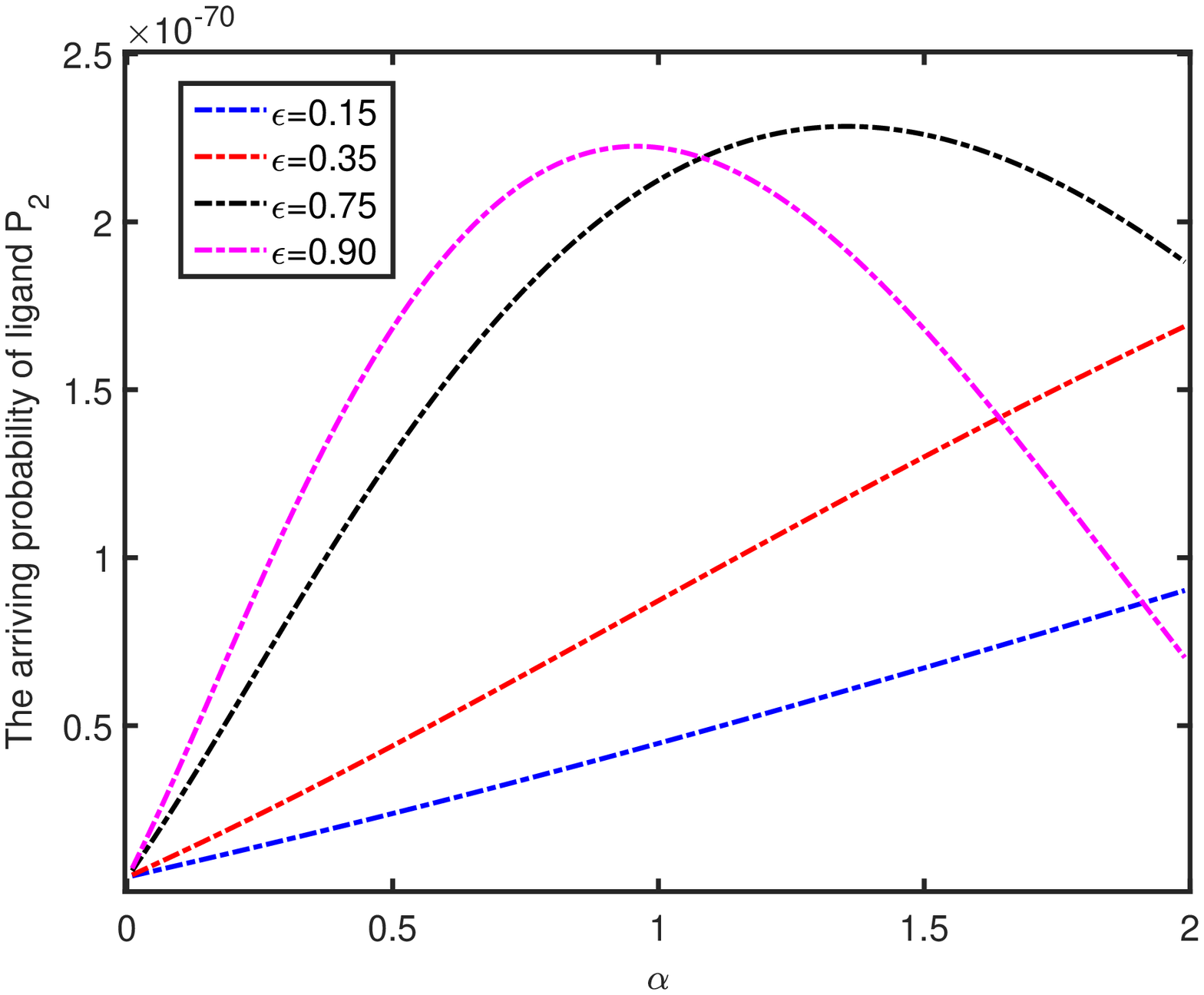}}
	\caption{\textbf{ Evolution with different non-Gaussian noise intensity $\epsilon$: T=5, $\sigma=0.15$.} (a) $[0,\pi]$. (b)$[\pi,2\pi]$.}
	\label{case3}
\end{figure}

In Figure \ref{case2}, the ligand locating probability $P_{2}$ is computed when  T=5 and $\epsilon=0.25$. We can see that in both $[0,\pi]$ and $[\pi,2\pi]$  domains, $P_{2}$ increases monotonically with respect to non-Gaussian index of L\'evy process $\alpha$. Besides, the noise intensity is helpful to increase the binding probability $P_{2}$.  We present the ligand locating probability $P_{2}$ with  T=5 and $\sigma=0.15$ in Figure \ref{case3}. For small L\'evy noise intensity, $P_{2}$ increases monotonically with respect to non-Gaussian index of L\'evy process $\alpha$; while for large  L\'evy noise intensity, $P_{2}$ increases first and then decreases with  $\alpha$. That points to the existence of an optimal  $\alpha$ which will lead to the max probability for the ligand to reach the binding site.

\begin{table}[!htb]
	\centering
	\small
	\caption{\ The probability $P_{2}$ on each segment with $\epsilon=0.25$, $\sigma=0.1$.}
	\label{tb10}
	\begin{tabular*}{1.0\textwidth}{@{\extracolsep{\fill}}llllllll}
		\hline
		$Segment $ & $(0, \frac{\pi}{4})$ & $(\frac{\pi}{4}, \frac{\pi}{2})$ &  $(\frac{\pi}{2}, \frac{3\pi}{4})$ &  $(\frac{3\pi}{4}, \pi)$ \\
		\hline
		$\alpha=0.5$ & $6.76\times10^{-187}$ &$9.46\times10^{-108}$ & $4.90\times10^{-50}$ & $9.36\times10^{-14}$ \\
		\hline
		$Segment $ &  $(\pi, \frac{5\pi}{4})$ &  $(\frac{5\pi}{4}, \frac{3\pi}{2})$  &  $(\frac{3\pi}{2}, \frac{7\pi}{4})$  &  $(\frac{7\pi}{4}, 2\pi)$ \\
		\hline
		$\alpha=0.5$ & $6.52\times10^{-298}$ & $2.42\times10^{-195}$ & $3.32\times10^{-114}$ &  $1.69\times10^{-54}$\\
		\hline\hline
		$Segment $ & $(0, \frac{\pi}{4})$ & $(\frac{\pi}{4}, \frac{\pi}{2})$ &  $(\frac{\pi}{2}, \frac{3\pi}{4})$ &  $(\frac{3\pi}{4}, \pi)$ \\
		\hline
		$\alpha=1.5$  & $1.26\times10^{-187}$ & $1.76\times10^{-108}$ & $9.13\times10^{-51}$  & $1.74\times10^{-14}$\\
		\hline
		$Segment $ &  $(\pi, \frac{5\pi}{4})$ &  $(\frac{5\pi}{4}, \frac{3\pi}{2})$  &  $(\frac{3\pi}{2}, \frac{7\pi}{4})$  &  $(\frac{7\pi}{4}, 2\pi)$ \\
		\hline
		$\alpha=1.5$  & $1.21\times10^{-298}$ & $4.51\times10^{-196}$ & $6.20\times10^{-115}$ &  $3.15\times10^{-55}$\\
		\hline
	\end{tabular*}
\end{table}

\begin{table}[!htb]
	\centering
	\small
	\caption{\ The probability $P_{2}$  on each segment with $\epsilon=0.25$, $\sigma=0.3$.}
	\label{tb11}
	\begin{tabular*}{1.0\textwidth}{@{\extracolsep{\fill}}llllllll}
		\hline
		$Segment $ & $(0, \frac{\pi}{4})$ & $(\frac{\pi}{4}, \frac{\pi}{2})$ &  $(\frac{\pi}{2}, \frac{3\pi}{4})$ &  $(\frac{3\pi}{4}, \pi)$ \\
		\hline
		$\alpha=0.5$ & $9.03\times10^{-187}$ &$1.26\times10^{-107}$ & $6.54\times10^{-50}$ & $1.25\times10^{-13}$ \\
		\hline
		$Segment $ &  $(\pi, \frac{5\pi}{4})$ &  $(\frac{5\pi}{4}, \frac{3\pi}{2})$  &  $(\frac{3\pi}{2}, \frac{7\pi}{4})$  &  $(\frac{7\pi}{4}, 2\pi)$ \\
		\hline
		$\alpha=0.5$ & $8.71\times10^{-298}$ & $3.23\times10^{-195}$ & $4.44\times10^{-114}$ &  $2.26\times10^{-54}$\\
		\hline\hline
		$Segment $ & $(0, \frac{\pi}{4})$ & $(\frac{\pi}{4}, \frac{\pi}{2})$ &  $(\frac{\pi}{2}, \frac{3\pi}{4})$ &  $(\frac{3\pi}{4}, \pi)$ \\
		\hline
		$\alpha=1.5$  & $1.39\times10^{-187}$ & $1.95\times10^{-108}$ & $1.01\times10^{-50}$  & $1.93\times10^{-14}$\\
		\hline
		$Segment $ &  $(\pi, \frac{5\pi}{4})$ &  $(\frac{5\pi}{4}, \frac{3\pi}{2})$  &  $(\frac{3\pi}{2}, \frac{7\pi}{4})$  &  $(\frac{7\pi}{4}, 2\pi)$ \\
		\hline
		$\alpha=1.5$  & $1.34\times10^{-298}$ & $4.99\times10^{-196}$ & $6.86\times10^{-115}$ &  $3.49\times10^{-55}$\\
		\hline
	\end{tabular*}
\end{table}

\begin{table}[!htb]
	\centering
	\small
	\caption{\ The probability $P_{2}$ on each segment with $\epsilon=0.4$, $\sigma=0.3$.}
	\label{tb12}
	\begin{tabular*}{1.0\textwidth}{@{\extracolsep{\fill}}llllllll}
		\hline
		$Segment $ & $(0, \frac{\pi}{4})$ & $(\frac{\pi}{4}, \frac{\pi}{2})$ &  $(\frac{\pi}{2}, \frac{3\pi}{4})$ &  $(\frac{3\pi}{4}, \pi)$ \\
		\hline
		$\alpha=0.5$  & $6.83\times10^{-187}$ &$9.56\times10^{-108}$ & $4.95\times10^{-50}$ & $9.45\times10^{-14}$ \\
		\hline
		$Segment $ &  $(\pi, \frac{5\pi}{4})$ &  $(\frac{5\pi}{4}, \frac{3\pi}{2})$  &  $(\frac{3\pi}{2}, \frac{7\pi}{4})$  &  $(\frac{7\pi}{4}, 2\pi)$ \\
		\hline
		$\alpha=0.5$ & $6.58\times10^{-298}$ & $2.45\times10^{-195}$ & $3.36\times10^{-114}$ &  $1.71\times10^{-54}$\\
		\hline\hline
		$Segment $ & $(0, \frac{\pi}{4})$ & $(\frac{\pi}{4}, \frac{\pi}{2})$ &  $(\frac{\pi}{2}, \frac{3\pi}{4})$ &  $(\frac{3\pi}{4}, \pi)$ \\
		\hline
		$\alpha=1.5$  & $6.43\times10^{-188}$ & $9.00\times10^{-109}$ & $4.66\times10^{-51}$  & $8.91\times10^{-15}$\\
		\hline
		$Segment $ &  $(\pi, \frac{5\pi}{4})$ &  $(\frac{5\pi}{4}, \frac{3\pi}{2})$  &  $(\frac{3\pi}{2}, \frac{7\pi}{4})$  &  $(\frac{7\pi}{4}, 2\pi)$ \\
		\hline
		$\alpha=1.5$ & $6.21\times10^{-299}$ & $2.30\times10^{-196}$ & $3.17\times10^{-115}$ &  $1.61\times10^{-55}$\\
		\hline
	\end{tabular*}
\end{table}

\section{The Most Probable Binding Probability}

The last section gives the quantitative illustration about the most probable probability of receptors and ligands location on the cell surface respectively. Based on that we could further get the most probable binding site on the surface by combining $P_{1}$ and $P_{2}$ and calculating the most probable binding probability $P=P_1P_2$. Indeed, it is hard to figure out the probabilities of all the possible parameters, so we just base on the probability of the partial location what we have figured out and provide the corresponding most probable binding probability.

The binding probability for receptors and ligands at different site with different parameters was shown in Tables 13-15.
We find that the binding probability is related with diffusion coefficient $\sigma$, non-Gaussnian noise coefficient $\epsilon$ and non-Gaussnian index $\alpha$. Note that $P_1$ has a common estimation factor, but we ignore it in the following calculations.
\begin{table}[!htb]
	\centering
	\small
	\caption{\ The binding probability $P$ on each segment with $\epsilon=0.25$, $\sigma=0.1$.}
	\label{tb13}
	\begin{tabular*}{1.0\textwidth}{@{\extracolsep{\fill}}llllllll}
		\hline
		$Segment $ & $(0, \frac{\pi}{4})$ & $(\frac{\pi}{4}, \frac{\pi}{2})$ &  $(\frac{\pi}{2}, \frac{3\pi}{4})$ &  $(\frac{3\pi}{4}, \pi)$ \\
		\hline
		$\alpha=0.5$ & $1.35\times10^{-188}$ & $2.85\times10^{-120}$ & $3.47\times10^{-57}$ & $5.68\times10^{-84}$ \\
		\hline
		$Segment $ &  $(\pi, \frac{5\pi}{4})$ &  $(\frac{5\pi}{4}, \frac{3\pi}{2})$  &  $(\frac{3\pi}{2}, \frac{7\pi}{4})$  &  $(\frac{7\pi}{4}, 2\pi)$ \\
		\hline
		$\alpha=0.5$ & $3.55\times10^{-323}$ & $3.09\times10^{-217}$ & $3.81\times10^{-203}$ &  \bm{$1.69\times10^{-56}$}\\
		\hline\hline
		$Segment $ & $(0, \frac{\pi}{4})$ & $(\frac{\pi}{4}, \frac{\pi}{2})$ &  $(\frac{\pi}{2}, \frac{3\pi}{4})$ &  $(\frac{3\pi}{4}, \pi)$ \\
		\hline
		$\alpha=1.5$ & $2.52\times10^{-189}$ & $5.31\times10^{-121}$ & $6.48\times10^{-58}$ & $1.05\times10^{-84}$ \\
		\hline
		$Segment $ &  $(\pi, \frac{5\pi}{4})$ &  $(\frac{5\pi}{4}, \frac{3\pi}{2})$  &  $(\frac{3\pi}{2}, \frac{7\pi}{4})$  &  $(\frac{7\pi}{4}, 2\pi)$ \\
		\hline
		$\alpha=1.5$ & $6.60\times10^{-324}$ & $5.77\times10^{-218}$ & $7.13\times10^{-204}$ &  \bm{$3.15\times10^{-57}$}\\
		\hline
	\end{tabular*}
\end{table}

\begin{table}[!htb]
	\centering
	\small
	\caption{\ The binding probability $P$ on each segment with $\epsilon=0.25$, $\sigma=0.3$.}
	\label{tb14}
	\begin{tabular*}{1.0\textwidth}{@{\extracolsep{\fill}}llllllll}
		\hline
		$Segment $ & $(0, \frac{\pi}{4})$ & $(\frac{\pi}{4}, \frac{\pi}{2})$ &  $(\frac{\pi}{2}, \frac{3\pi}{4})$ &  $(\frac{3\pi}{4}, \pi)$ \\
		\hline
		$\alpha=0.5$  & $1.80\times10^{-188}$ & $3.80\times10^{-120}$ & $4.64\times10^{-57}$ & $7.58\times10^{-84}$ \\
		\hline
		$Segment $ &  $(\pi, \frac{5\pi}{4})$ &  $(\frac{5\pi}{4}, \frac{3\pi}{2})$  &  $(\frac{3\pi}{2}, \frac{7\pi}{4})$  &  $(\frac{7\pi}{4}, 2\pi)$ \\
		\hline
		$\alpha=0.5$  & $4.75\times10^{-323}$ & $4.13\times10^{-217}$ & $5.10\times10^{-203}$ &  \bm{$2.26\times10^{-56}$}\\
		\hline\hline
		$Segment $ & $(0, \frac{\pi}{4})$ & $(\frac{\pi}{4}, \frac{\pi}{2})$ &  $(\frac{\pi}{2}, \frac{3\pi}{4})$ &  $(\frac{3\pi}{4}, \pi)$ \\
		\hline
		$\alpha=1.5$ & $2.78\times10^{-189}$ & $5.88\times10^{-121}$ & $7.17\times10^{-58}$ & $1.17\times10^{-84}$ \\
		\hline
		$Segment $ &  $(\pi, \frac{5\pi}{4})$ &  $(\frac{5\pi}{4}, \frac{3\pi}{2})$  &  $(\frac{3\pi}{2}, \frac{7\pi}{4})$  &  $(\frac{7\pi}{4}, 2\pi)$ \\
		\hline
		$\alpha=1.5$ & $7.31\times10^{-324}$ & $6.38\times10^{-218}$ & $7.88\times10^{-204}$ &  \bm{$3.49\times10^{-57}$}\\
		\hline
	\end{tabular*}
\end{table}

\begin{table}[!htb]
	\centering
	\small
	\caption{\ The binding probability $P$ on each segment with $\epsilon=0.4$, $\sigma=0.3$.}
	\label{tb15}
	\begin{tabular*}{1.0\textwidth}{@{\extracolsep{\fill}}llllllll}
		\hline
		$Segment $ & $(0, \frac{\pi}{4})$ & $(\frac{\pi}{4}, \frac{\pi}{2})$ &  $(\frac{\pi}{2}, \frac{3\pi}{4})$ &  $(\frac{3\pi}{4}, \pi)$ \\
		\hline
		$\alpha=0.5$ & $1.36\times10^{-188}$ & $2.88\times10^{-120}$ & $3.51\times10^{-57}$ & $5.73\times10^{-84}$ \\
		\hline
		$Segment $ &  $(\pi, \frac{5\pi}{4})$ &  $(\frac{5\pi}{4}, \frac{3\pi}{2})$  &  $(\frac{3\pi}{2}, \frac{7\pi}{4})$  &  $(\frac{7\pi}{4}, 2\pi)$ \\
		\hline
		$\alpha=0.5$ & $3.59\times10^{-323}$ & $3.06\times10^{-217}$ & $3.86\times10^{-203}$ &  \bm{$1.71\times10^{-56}$}\\
		\hline\hline
		$Segment $ & $(0, \frac{\pi}{4})$ & $(\frac{\pi}{4}, \frac{\pi}{2})$ &  $(\frac{\pi}{2}, \frac{3\pi}{4})$ &  $(\frac{3\pi}{4}, \pi)$ \\
		\hline
		$\alpha=1.5$ & $1.28\times10^{-189}$ & $2.17\times10^{-121}$ & $3.30\times10^{-58}$ & $5.40\times10^{-85}$ \\
		\hline
		$Segment $ &  $(\pi, \frac{5\pi}{4})$ &  $(\frac{5\pi}{4}, \frac{3\pi}{2})$  &  $(\frac{3\pi}{2}, \frac{7\pi}{4})$  &  $(\frac{7\pi}{4}, 2\pi)$ \\
		\hline
		$\alpha=1.5$ & $3.39\times10^{-324}$ & $2.94\times10^{-218}$ & $3.64\times10^{-204}$ &  \bm{$1.61\times10^{-57}$}\\
		\hline
	\end{tabular*}
\end{table}

We have picked out the  most probable binding location corresponding to the largest binding probability in each case in bold font. According to the  results, we find that the  most probable binding location with largest probability is in $(\frac{7\pi}{4}, 2\pi)$,  both in $\alpha=0.5$ and $\alpha=1.5$ cases, with our choosen noise intensities $\sigma$ and $\epsilon$. In this way, we can make  predictions in the cases of other non-Gaussian index $\alpha$ and noise intensities $\sigma$ and $\epsilon$. This receptor-ligand binding prediction method is based on a simple diffusion model for the receptor and a jump-diffusion model for the ligand.

\section{Conclusion}
In this paper, we have considered stochastic dynamics for an cell surface receptor binding to an extracellular ligand on the cell membrane. The receptor is modeled as a usual diffusion, while the ligand is regarded as a pure jump-diffusion. To explore the binding probability, we divided the cell surface into eight segments, and calculate the largest probabilities of receptor and ligand reaching each segment.
As for the receptor, we  used the Onsager-Machlup functional to compute the most probable transition pathway from the cell center to the cell membrane. To solve the Euler-Lagrange equation, we proposed a neural shooting method for the most probable transition pathway. We further on obtained the action functional value as well as the transition probability.
The motion of extracellular ligand is quantized by a fractional  Fokker-Planck equation. We used a fast discretization method to solve this equation and got the probability of the ligand arriving at each segment on the membrane. We also investigated the influence of the system parameters on the ligand arriving probability on cell membrane, and found the Brownian noise had advantage on locating the
closer site on the membrane over L\'evy noise, while the phenomenon flips when the ligand arrived at a distant site. This result is consistent with that discussed in \cite{Palyulin2014,Chen2019}.

Based on the probabilities of the receptor diffusing to the cell membrane and the ligand arriving at the cell membrane, we further compute the largest binding probability and indicate the most probable binding site. In our study, the binding event happens once the receptor and ligand arrive at the same location on the cell membrane, which is a rare event. Our investigation provides a quantitative analysis for the extreme statistics. Besides, our work helps understand the receptor-ligand binding on the membrane, which also sheds light on                understanding of cell's response to external stimuli and communication with other cells.

To sum up, we have devised a method for predicting the likelihood for receptor-ligand binding on cell surface.
Specifically,

(a)We use neural networks based on Onsager-Machlup function to compute the most probable transition pathway of the receptor diffusion to the cell membrane.

(b)We compute the probability of the receptor diffusion to a site on the cell membrane and the probability of ligand locating  the same site respectively. We also compare the impact of non-Gaussnian noise with Gaussnian noise in the ligand dynamics.

(c)We combine the probability of the receptor diffusion to a site on  the cell membrane and the probability of ligand locating the same site to obtain the most probable binding site.

\section*{Acknowledgments}

We would like to thank Jianyu Chen for helpful discussions. This work is supported by the National Natural Science Foundation of China (NSFC) (Grant No.11901536 and No.12101473 ).

\section*{Data Availability}
 The data that support the findings of this study are openly available in Github {https://github.com/StringsLi/binding-code}.

 \section*{Appendix}

\subsection*{ L\'evy motion }
A scalar stable L\'evy motion $L_t^{\alpha}$, for $0<\alpha<2$,  is a non-Gaussian stochastic process with the following properties \cite{Duan,Applebaum2009,Sato1999,Samorodnitsky1994}: \\
(i)  $L_0^{\alpha} = 0$, almost surely;\\
(ii) $L_t^{\alpha}$ has independent increments;\\
(iii)$L_t^{\alpha}$ has stationary increments: $L_t^{\alpha}-L_s^{\alpha} $ has probability distribution $S_\alpha((t-s)^\frac{1}{\alpha}, 0, 0)$ for   $  s \leq t  $; in particular,  $L_t^{\alpha}$ has distribution $S_\alpha(t^\frac{1}{\alpha}, 0, 0)$; \\
(iv) $L_t^{\alpha}$ has stochastically continuous sample paths, i.e.,   $L_t^{\alpha} \rightarrow L_s^{\alpha}$ in probability, as $t\rightarrow s$.

Here $S_{\alpha}(\sigma,\beta,\mu)$ is the so-called stable distribution  \cite{Duan,Samorodnitsky1994} and is  determined by four parameters, non-Gaussianity index
 $\alpha  (0 < \alpha < 2)$, skewness index $\beta (-1\leq \beta \leq 1)$, shift index $\mu  (-\infty < \mu < +\infty)$ and scale index $\sigma  (\sigma \geq 0)$.

 The stable L\'evy motion $L_t^\alpha$ has  the jump measure
 $$
\nu_{\alpha}(dy)=C_\alpha |y|^{-(1+\alpha)}\, dy,
$$
 where the coefficient
 $$
C_{\alpha} =
\frac{\alpha}{2^{1-\alpha}\sqrt{\pi}}
\frac{\Gamma(\frac{1+\alpha}{2})}{\Gamma(1-\frac{\alpha}{2})}.
$$


Note that the well-known Brownian motion $B_t$  is a special case corresponding to $\alpha=2$.

\subsection*{Nonlocal Fokker-Planck equation and numerical methods}

Consider a scalar stochastic differential equation with L\'evy noise
\begin{equation} \label{sde}
  d X_t = f(X_t) dt + \varepsilon d L_t,   \;\; X_0= x_0,
\end{equation}
where $f$ is a given   vector field (or drift) and $ \varepsilon$ is the noise intensity.

The generator for this  stochastic differential equation is
\begin{equation}
A\varphi(x)=f(x)\varphi'(x)  +  \int_{\mathbb{R}^{1}\backslash \{0\}}[\varphi(x + \varepsilon y) - \varphi(x)] \nu_\alpha(dy).    \label{gener1}
\end{equation}
Let $ z= \varepsilon y$.  The generator becomes
\begin{equation*}
A\varphi(x)=f(x)\varphi'(x)  + \varepsilon^\alpha  \int_{\mathbb{R}^{1}\backslash \{0\}}[\varphi(x + z) - \varphi(x)]\nu_\alpha(dz).  \label{gener2}
\end{equation*}

The Fokker-Planck  equation for this stochastic differential equation, i.e., the probability
density $p(x,t)$ for the solution process $X_t $ with initial condition $X_0=x_0$ is \cite{Duan}
\begin{equation} \label{fpe}
 p_t  = A^* p,  \;\;   p(x,0)=\delta(x-x_0),
\end{equation}
where  $A^*$ is the adjoint operator of the generator  $A$  in   Hilbert space $ L^2(R^1) $, as defined by
\begin{equation*}
\int_{\mathbb{R}^{1}\backslash \{0\}} A\varphi(x)u(x)dx = \int_{\mathbb{R}^{1}\backslash \{0\}}\varphi(x)A^*u(x)dx.
\end{equation*}
Then via integration by parts, we get the adjoint operator for $A$
\begin{equation}
A^*u(x)=\int_{\mathbb{R}^{1}\backslash \{0\}} \varepsilon^\alpha [u(x-y)- u(x)] \; \nu_\alpha(dy).  \label{hilbert}
\end{equation}
Thus we have the nonlocal Fokker-Planck equation
\begin{equation} \label{fpe2}
p_t = - (f(x)p(x, t))_x +\int_{\mathbb{R}^{1}\backslash \{0\}} \varepsilon^\alpha [p(x-y, t)- p(x, t)] \; \nu_\alpha(dy).
\end{equation}

We use a    numerical finite difference method developed  in Gao et al. \cite{Gao2016} to simulate the nonlocal Fokker-Planck  equation  (\ref{fpe2}).

\end{document}